\documentclass[
    ,final            
  ]
  {aipproc}

\layoutstyle{6x9}

\begin{document}

\title{GD 244: asteroseismology of a pulsator in the middle of the ZZ Ceti instability strip}

\classification{97.10.Cv, 97.10.Sj, 97.20.Rp, 97.30.Dg}
\keywords      {asteroseismology, stellar pulsations, white dwarfs, star: GD 244}

\author{Zs. Bogn\'ar}{
  address={Konkoly Observatory of the Hungarian Academy of Sciences, P.O. Box 67., H--1525 Budapest, Hungary}
}

\author{M. Papar\'o}{
  address={Konkoly Observatory of the Hungarian Academy of Sciences, P.O. Box 67., H--1525 Budapest, Hungary}
}

\begin{abstract}
We present our preliminary results on the asteroseismological investigations of the ZZ Ceti star GD 244. 
We used literature values of the effective temperature and surface gravity and utilized the White Dwarf 
Evolution Code of Bischoff-Kim, Montgomery and Winget (2008, ApJ, 675, 1512) to build our model grid for 
the seismological analysis. Five observed pulsational modes published up to now were used to find acceptable 
model solutions. We found that the best model fits have masses between 0.61 and 0.74\,$M_{\odot}$ and constitute 
two groups with hydrogen layer masses of either $\sim10^{-5}$ or $10^{-6}\,M_{\odot}$. Based on a statistical 
analysis of a larger sample of possible model solutions, we assume that the mass of the star is below 
$\sim0.68\,M_{\odot}$ and the oxygen content in the centre is less than 60 per cent.
\end{abstract}

\maketitle


\section{Initial parameters and the model grid}

Our aim is to find stellar models that match the observed properties of the ZZ Ceti star GD 244 with high 
precision. We present our asteroseismological results using published period and atmospheric parameters.

GD 244 was observed with the Canada-France-Hawaii Telescope (CFHT) in 1999 (\mbox{1 night}, \cite{fontaine1}) 
and at McDonald Observatory in 2003 (10 nights, \cite{yeates1}).  Table~\ref{periods} shows the period and 
amplitude values detected and the assumption about mode identification. Not all the periods were observed in 
both data sets. Amplitude variations may be responsible for the differences. In such cases we could attempt 
to find normal modes enough for asteroseismology by taking all known modes from the different seasons (see e.g. 
\cite{kleinman1}). Period values used for the seismological analysis are typeset in boldface. All but one were 
determined from the McDonald Observatory data set, which had better frequency resolution than the short CFHT data. 
In three cases the same $m =$ -1 value was presumed for the modes. 

\begin{table}
\begin{tabular}{lcccrlr}
\hline
 & \multicolumn{2}{c}{CFHT 1999} & \multicolumn{4}{c}{McDO 2003} \\
 & $P$ [s] & $A$ [No.] & $P$ [s] & $A$ [mma] & $l$ & $m$\\
\hline
 $f_\mathrm{1}$& 203.3 & 4.& \bf{202.98} & 4.04 & 1 & -1?\\
 $f_\mathrm{2}$& 256.3 & 2. & \bf{256.56} & 12.31 & 1? & -1?\\
 & & & 256.20 & 6.73 & 1? & +1?\\
 $f_\mathrm{3}$& \bf{294.6} & 3.& & &\\
 $f_\mathrm{4}$& 307.0 & 1.& \bf{307.13} & 20.18 & 1 & -1\\
 & & & 306.57 & 5.02 & 1 & +1\\
 $f_\mathrm{5}$& & & \bf{906.08} & 1.72 & $\le$3 & ?\\ 
\hline
\end{tabular}
\caption{Pulsation modes of GD 244. $A$ [No.] refers to relative values estimated by Fig.\,3 in \cite{fontaine1}.}
\label{periods}
\end{table}

\begin{table}
\begin{tabular}{lcl}
\hline
$T_{\mathrm{eff}}$ [K] & log\,$g$ [cgs] & References \\
\hline
11\,680 & 8.08 & Fontaine et al. \cite{fontaine1} \\
11\,611 & 7.91 & Koester et al. \cite{koester1} \\
11\,293 & 8.21 & \\
11\,640 & 8.05 & Limoges \& Bergeron \cite{limoges1} \\
\hline
\end{tabular}
\caption{Effective temperature and surface gravity values determined by spectroscopy.}
\label{athm}
\end{table}

Table~\ref{athm} summarizes the atmospheric parameters determined by optical spectroscopy. We covered the 
log\,$g$ range and a wider range in $T_{\mathrm{eff}}$ with our models. The latter approximately corresponds 
to the interval of ZZ Ceti stars.

We used the White Dwarf Evolution code of Bischoff-Kim, Montgomery and Winget \cite{bkim1} to build our model grid. 
We varied five input parameters of the WDEC in the following ranges (in square brackets: step sizes):\\
\newpage
\noindent$T_{\mathrm{eff}} =$ 10\,800 -- 12\,200 [200]\,K,\\
$ M_* =$ 0.525 -- 0.740 [0.005]\,$M_\odot$ (log\,$g \sim$ 7.85 -- 8.23),\\
$M_{\mathrm{H}} = 10^{-4} - 10^{-8}$ [$10^{-0.2}$]\,$M_*$,\\
$X_{\mathrm{O}} =$ 0.5 -- 0.9 [0.1] (central oxygen abundance),\\
$X_{\mathrm{fm}} =$ 0.1 -- 0.5 [0.1] (the fractional mass point where the oxygen abundance starts dropping).\\
We fixed the mass of the helium layer at $10^{-2}\,M_*$.

\section{Results and discussion}

Our first criterion for an acceptable model was that it should give period values close to the observed ones. 
We used the parameter root main square (\emph{r.m.s.}) calculated from the observed and model periods to select 
among the possible solutions. The fitting routine \textsc{fitper} \cite{kim1} was applied for this purpose. 
Since we had only five modes to fit, many acceptable models had low \emph{r.m.s.} value. We let all five modes to 
be $l =$ 1 or 2 for the fitting procedure. Assuming better visibility of $l =$ 1 modes, we selected the models that 
gave at least three $l =$ 1 solutions. Beside the low \emph{r.m.s.}, this was our second criterion.

Table~\ref{models} summarizes the parameters of our best-fitting models. The last column shows the $l =$ 2 modes only. 
We can discriminate between the models based on the observed amplitudes. Considering that the 256 and 307-s modes have 
the largest amplitudes, it is improbable that both of them are $l =$ 2. Thus, the 0.665 and 0.685\,$M_\odot$ models 
(both at the low temperature region) are less favourable. The 0.620\,$M_\odot$ model represents an interesting case. 
It has the thinnest hydrogen layer and only this gives $l =$ 1 value for both the dominant modes. The rest of the models 
form two groups: one with $10^{-6}\,M_*$ hydrogen layers and $l =$ 2 values for the 256 and 294-s modes, and another with 
$\sim 10^{-5}\,M_*$ hydrogen layers and $l =$ 2 value for the 307-s mode. Since the 307-s mode has the largest light amplitude, 
being $l =$ 2 would imply a much larger physical amplitude, thus we prefer an $l =$ 1 solution for this mode. Therefore, the 
models with stellar masses between 0.610 -- 0.630\,$M_\odot$ and $M_{\mathrm{H}} = 10^{-6}\,M_*$ or the one with 
$M_{\mathrm{H}} = 1.6\,\times\,10^{-7}\,M_*$ are better choices. The 203-s mode is always $l =$ 1.

Yeates et al. \cite{yeates1} also identified the 203-s mode as $l =$ 1 using combination frequency amplitudes. 
Castanheira \& Kepler \cite{castanheira1} gave $l =$ 2 value for this mode. They built their model grid using fixed, 
homogeneous C/O 50:50 cores, but varied the mass of the helium layer. Only the 203, 256 and 307-s modes were fitted. 
Their best solution was $T_{\mathrm{eff}} =$ 12\,200\,K, $M_* =$ 0.68\,$M_\odot$, $M_{\mathrm{H}} = 10^{-7}\,M_*$ and 
$M_{\mathrm{He}} = 10^{-3.5}\,M_*$.

\begin{table}
\begin{tabular}{lcccccr}
\hline
\multicolumn{1}{l}{$M_*$/$M_{\odot}$} & \multicolumn{1}{c}{$T_{\mathrm{eff}}$\,[K]} & \multicolumn{1}{c}{-log\, $M_\mathrm{H}$} 
& \multicolumn{1}{c}{$X_\mathrm{O}$} & \multicolumn{1}{c}{$X_{\mathrm{fm}}$} & \multicolumn{1}{c}{\emph{r.m.s.}\,[s]} 
& \multicolumn{1}{c}{$l =$ 2 [s]}\\
\hline
0.610 & 12\,000 & 6.0 & 60 & 50 & 0.95 & $f_\mathrm{2}$, $f_\mathrm{3}$\\
0.615 & 11\,800 & 6.0 & 50 & 40 & 0.91 & $f_\mathrm{2}$, $f_\mathrm{3}$\\
0.615 & 11\,800 & 6.0 & 70 & 50 & 0.89 & $f_\mathrm{2}$, $f_\mathrm{3}$\\
0.620 & 11\,600 & 6.8 & 80 & 50 & 0.88 & $f_\mathrm{3}$, $f_\mathrm{5}$\\
0.625 & 12\,200 & 5.0 & 50 & 10 & 0.97 & $f_\mathrm{4}$\\
0.630 & 12\,000 & 5.0 & 50 & 10 & 0.91 & $f_\mathrm{4}$\\
0.630 & 11\,400 & 6.0 & 70 & 50 & 0.69 & $f_\mathrm{2}$, $f_\mathrm{3}$\\
0.640 & 11\,800 & 5.0 & 50 & 10 & 0.84 & $f_\mathrm{4}$\\
0.665 & 10\,800 & 5.0 & 80 & 30 & 0.67 & $f_\mathrm{2}$, $f_\mathrm{4}$\\
0.685 & 10\,800 & 5.2 & 60 & 20 & 1.17 & $f_\mathrm{2}$, $f_\mathrm{4}$\\
0.730 & 11\,600 & 4.8 & 80 & 20 & 1.12 & $f_\mathrm{3}$, $f_\mathrm{4}$\\
0.735 & 11\,400 & 4.8 & 80 & 20 & 0.98 & $f_\mathrm{3}$, $f_\mathrm{4}$\\
\hline
\end{tabular}
\caption{Parameters of our best-fitting models. We indicate the \emph{r.m.s.} values calculated from the 
observed and model periods in column 6. The last column shows the modes of Table~\ref{periods} that are $l =$ 2 
according to the model. The other modes are $l =$ 1.}
\label{models}
\end{table}

\begin{figure}
  \includegraphics[]{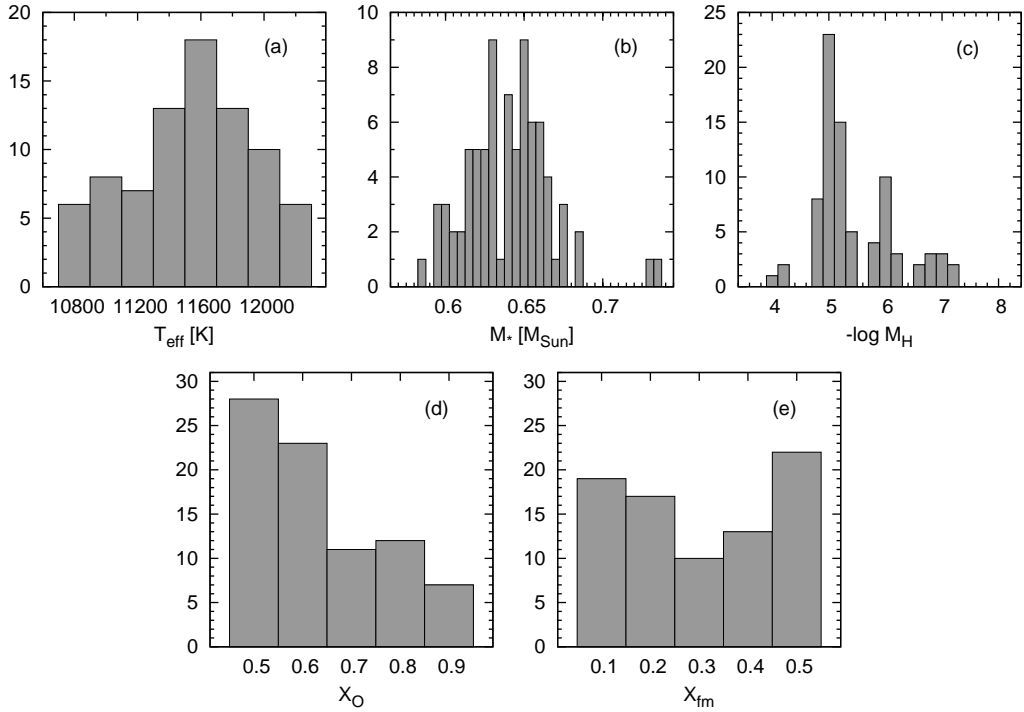}
  \caption{Histograms of the 81 selected models for the five physical parameters varied in the grid. The bin sizes 
  are suited to the grid step sizes.}
  \label{hist}
\end{figure}

We investigated the acceptable models from a statistical point of view also. We extended our sample selecting 
every model that gave at least three $l =$ 1 solutions and had \emph{r.m.s.} $<$ 1.5 values. With these criteria 
we obtained significantly more, 81 models for the analysis. The histograms in Fig.~\ref{hist} show the number of 
models in the bins given by the step sizes for the five physical parameters.

As Fig.\,1a shows, most of the solutions are in the 11\,400 -- 12\,000\,K $T_{\mathrm{eff}}$ range and the most 
frequent value is 11\,600\,K. Most of the spectroscopic $T_{\mathrm{eff}}$ values are also around 11\,600\,K. 
Considering Fig.\,1b, the populated bins can be found between 0.62 and 0.67\,$M_\odot$. Peak values are at 0.63 
and 0.65\,$M_\odot$. Fig.\,1c clearly shows the two families of solutions with $M_\mathrm{H} \sim 10^{-5}$ and 
$10^{-6}\,M_*$. The favoured hydrogen layer mass with this method is $10^{-5}\,M_*$. According to Fig.\,1d, the 
central oxygen abundance could be 50 -- 60\% (or less), larger values are not preferred. We can not give such 
constraint on the $X_{\mathrm{fm}}$ parameter (Fig.\,1e). 

Additional photometric observations were obtained on this star both at McDonald Observatory and Piszk\'estet\H o 
mountain station of Konkoly Observatory. The analysis of these data and a detailed asteroseismological investigation 
will be the subject of a forthcoming paper.   

%



\bibliographystyle{aipprocl} 





\end{document}